# Deep learning can differentiate IDH-mutant from IDH-wild type GBM

**Luca Pasquini**[1,2*], **Antonio Napolitano**[3*], **Emanuela Tagliente**[3], **Francesco Dellepiane**[1], **Martina Lucignani**[3], **Antonello Vidiri**[4], **Giulio Ranazzi**[5], **Antonella Stoppacciaro**[5], **Giulia Moltoni**[1], **Matteo Nicolai**[1], **Andrea Romano**[1], **Alberto Di Napoli**[1], **Alessandro Bozzao**[1]

[1]Neuroradiology Unit, NESMOS Department, Sant'Andrea Hospital, La Sapienza University, Via di Grottarossa 1035, Rome 00189, Italy

[2]Neuroradiology Unit, Radiology Department, Memorial Sloan Kettering Cancer Center, 1275 York Ave, New York, NY 10065, USA

[3]Medical Physics Department, Bambino Gesù Children's Hospital, IRCCS, Piazza di Sant'Onofrio, 4, Rome 00165, Italy

[4]Radiology and Diagnostic Imaging Department, Regina Elena National Cancer Institute, IRCCS, Via Elio Chianesi 53, Rome 00144, Italy

[5]Department of Clinical and Molecular Medicine, Surgical Pathology Unit, Sant'Andrea Hospital, La Sapienza University, Via di Grottarossa 1035, Rome 00189, Italy

* Authors contributed equally to this work.

Correspondence to:
Luca Pasquini
Neuroradiology Unit, NESMOS Department, Sant'Andrea Hospital, La Sapienza University, Via di Grottarossa 1035,
Rome 00189, Italy
Email: lucapasquini3@gmail.com
Phone: 06.3377.5225
ORDIC: 0000-0003-1679-6958

**Keywords:** GBM, MRI, IDH, deep-learning, CBV, artificial intelligence

**Background**: Distinction of IDH mutant and wildtype GBMs is challenging on MRI, since conventional imaging shows considerable overlap. While few studies employed deep-learning in a mixed low/high grade glioma population, a GBM-specific model is still lacking in the literature.

Our objective was to develop a deep-learning model for IDH prediction in GBM by using Convoluted Neural Networks (CNN) on multiparametric MRI.

**Methods**: We included 100 adult patients with pathologically proven GBM and IDH testing. MRI data included: morphologic sequences, rCBV and ADC maps. Tumor area was obtained by a bounding box function on the axial slice with widest tumor extension on T2 images and was projected on every sequence. Data was split into training and test (80:20) sets. A 4 block 2D - CNN architecture was implemented for IDH prediction on every MRI sequence. IDH mutation probability was calculated with softmax activation function from the last dense layer. Highest performance was calculated accounting for model accuracy and categorical cross-entropy loss (CCEL) in the test cohort.

**Results**: Our model achieved the following performance: T1 (accuracy 77%, CCEL 1.4), T2 (accuracy 67%, CCEL 2.41), FLAIR (accuracy 77%, CCEL 1.98), MPRAGE (accuracy 66%, CCEL 2.55), rCBV (accuracy 83%, CCEL 0.64). ADC achieved lower performance.

**Conclusion**: We built a GBM-tailored deep-learning model for IDH mutation prediction, achieving accuracy of 83% with rCBV maps. High predictivity of perfusion images may reflect the known correlation between IDH, hypoxia inducible factor (HIF) and neoangiogenesis. This model may set a path for non-invasive evaluation of IDH mutation in GBM.

INTRODUCTION

Glioblastoma Multiforme (GBM) is the most lethal primary brain tumor of the adult and accounts for 70-75% of all diffuse gliomas [1] and 16% of primary CNS tumors in adults [2]. Despite state-of-the-art treatment implementing the combination of temozolomide and tailored radiotherapy, overall survival is very limited, with a median between 14 and 17 months [1]. Genetic profile directly impacts diagnosis and therapy of gliomas, with demonstrated effects on survival [3]. One of the most important genetic biomarkers of GBM is isocitrate dehydrogenase (IDH) [4]. Most GBMs are primary (90%) and rarely harbor IDH mutation (3.7%) [4]. Secondary GBMs represent around 10% of the total cases and are more likely to be mutated (73%) [4]. Overall, IDH mutation is expected in about 10% of GBMs [5]. Most important, mutant GBMs are characterized by significantly improved survival than wild-type GBMs (31 months vs. 15 months)

[3,4,6-8]. Also, patients with IDH1-mutated glioblastomas show better outcome than IDH1 wild-type gliomas of lower grade [9].

The gold-standard procedure for diagnosis of GBM is pathological sampling through brain biopsy or surgery. Along with the risk for complications, high costs and misinterpretation rate, biopsy-based methods may face incomplete sampling due to spatial heterogeneity of GBMs, characterized by multiple intra-tumoral habitats and variable genetic expression [10,11]. Also, various studies reported limitations for IDH mutation pathological testing, due to technical shortcomings [12-14], supporting the need for non-invasive diagnostic procedures to complement the ground truth of pathologic evidence. Magnetic Resonance Imaging (MRI) is a versatile examination that can provide noninvasive definition of genetic alterations in the pre-operative setting, guide targeted biopsies, screen highly probable genetic mutations for further testing and help tailoring treatment interventions to the single case [15].

Despite the importance for patient outcome and treatment, IDH-mutation does not present a clear radiologic signature [5]. For example, IDH mutant gliomas may show less enhancement, less blood flow on perfusion-weighted images, higher mean diffusion values, smaller size and frontal lobe location [16]. However, the sensitivity of these findings is somehow disappointing in GBM [5,17] (**Fig.1**).

In the last few years, the implementation of artificial intelligence in the field of radiomics held a great advancement in our understanding of the correlations between radiological features and genetic phenotypes [15]. Supervised machine learning (ML) techniques achieved good performance in predicting IDH mutation in GBM [16]. However, these techniques are time consuming and require expert supervision, being far from clinical implementation. Deep-learning may offer a solution to these shortcomings, thanks to its unsupervised nature and the ability to independently learn which features are most relevant for the task, without the constraint of 'a priori' feature selection [18]. Previous studies based on this approach mostly focused on differentiating IDH status in heterogeneous papulations of low grade gliomas (LGG) and high grade gliomas (HGG) [18]. A GBM-specific model for IDH prediction is lacking evidence in the literature, despite being highly relevant due to paucity of non-invasive biomarkers.

Our objective was to build a reliable deep-learning model predictive of IDH mutation in patients with GBM, starting from conventional and advanced MRI data.

## MATERIALS AND METHODS

*Subjects*

This retrospective observational study was conducted in agreement with the Helsinki declaration and was approved by the IRB (protocol number: 19 SA_2020). We enrolled patients who underwent preoperative MRI from March 2005 to May 2019, fulfilling the following inclusion criteria: histopathological diagnosis of GBM, MRI acquisition in the preoperative phase with at least one among structural, diffusion or perfusion techniques, IDH testing. Exclusion criteria were motion artifacts or other causes of sub-optimal images.

*Histopathological Analysis*

The specimens were fixed in 10% formaldehyde and processed for paraffin embedding. Two µm thick sections were mounted and stained with hematoxylin and eosin. Histopathological examination, typing and grading were performed identifying at least three of the following features in astrocytic tumors: mitotic figures, cellular atypia, microvascular proliferation and/or necrosis, according to the last edition of the World Health Organization classification of CNS tumors.

Immunohistochemistry was performed using Dako Envision Flex system. For the evaluation of IDH-1 mutation, IDH-1 R132H antibody was used. The test was defined as positive if a focal or diffuse immunopositivity was detected and negative if no tumor cells showed immunopositivity. Negative cases were then analyzed for IDH-1/2 mutations by directly sequencing the exon 4 of *IDH1* gene including codon 132, and a fragment of 219 bp in length spanning the catalytic domain of *IDH2* including codon 172 following PCR amplification of genomic DNA using respectively the primers *IDH-1*: Forward, 5′-CGG TCT TCA GAG AAG CCA TT-3′, Reverse, 5′- ATT CTT ATC TTT TGG TAT CTA CAC C-3′, *IDH-2*: forward 5'-CAAGCTGAAGAAGATGTGGAA-3′, reverse 5′ CAGAGACAAGAGGATGGCTA-3′. All sequence reactions were carried out using the GenomeLab DTCS quick-start kit (Beckman Coulter, Fullerton, CA, USA). The reactions were carried out in an automated DNA analyzer (CEQ 8000; Beckman Coulter).

*MR image acquisition*

Acquired MR sequences included MPRAGE, FLAIR, T1-weighted, T2-weigthed, diffusion weighted images (DWI), with apparent diffusion coefficient (ADC) map reconstruction, and perfusion weighted images (PWI) with dynamic susceptibility contrast (DSC) technique. MRI examinations were acquired on a 1.5T (Magnetom Sonata, Siemens, Erlangen, Germany) and a 3T scanner (Discovery MR750w; GE Medical Systems, Waukesha, WI, USA).

Patients underwent the following protocol: axial T1-weighted spin echo, axial T2-weighted fast spin echo, axial FLAIR; DWI acquired with three levels of diffusion sensitization (b-values 0, 500 and 1000); DSC acquired during contrast injection (DOTAREM.; dose 0.1 mmol/kg, injection rate 4 ml/s), followed by a 20-ml saline flush, based on T2*- weighted gradient-echo echo-planar; MPRAGE after administration of contrast. Perfusion parametric maps were obtained through a dedicated software package OleaSphere software version 3.0 (Olea Medical, La Ciotat, France). A rCBV map was generated by using an established tracer kinetic model applied to the first-pass data [19]. As previously described [20], the dynamic curves were mathematically corrected to reduce contrast agent leakage effects.

*Image processing*

All MR images were co-registered to individual MPRAGE sequence. In order to normalize pixels intensity, the tumor lesion area was derived by a bounding box function implemented on Python3. Data was masked according to a region of interest (ROI) obtained from T2 images in order to obtain the pixels of the whole tumor region, including peritumoral edema. All the others pixels were set to zero and the mean of the tumor region was used for scaling the entire volume data. Images intensities were normalized by subtracting the median intensity of the entire brain and dividing by the standard deviation of the same portion, as used in previous work [21]. For each patient, the axial slice with the widest tumor extension was selected, and a bounding rectangle derived from the tumor mask was drawn around the tumor. As each tumor was different in size, all images were resampled to 64X64 [22,23].

As the dataset under investigation included GBM patients only, IDH labels were unbalanced towards wildtype tumors [4]; for this reason, a threshold of at least 5 IDH-mutant patients was set for the testing group during randomization. To minimize overfitting, data augmentation techniques were employed. Once the dataset was split into training and testing sets (test_size=20%), augmentation allowed to fix the unbalance between classes in the training set

[23-25]. The augmentation method Image Data Generator class (Keras API) was employed with rotation_range = [0°,90°]. Other geometric transformations influencing tumor shape (such as flipping, color space, cropping, translation, noise injection) [26] were excluded as they might interfere with the final prediction accuracy, since shape features correlate to IDH mutation in our experience [unpublished data].

*CNN Architecture*

We propose a 2D CNN model with a set of 2D trainable filters. CNN derives high-level features from the low-level input, while estimated high-level features directly contribute to the classification of input data. The network architecture usually consists of a number of layers, which generate progressively higher-level features as we go deeper into the network. Inspired by very deep convolutional networks (e.g. VGGNet, ResNet), we designed a 4 blocks 2D - CNN architecture (**Fig.2**). Rectified linear layer (ReLU) and batch normalization, both after the 2D convolutional layer, compose the convolutional block. The CNN input was a 64x64 2D MR slice extracted at the level of maximal tumor extension. The convolutional layers computed their outputs from the input slice by applying convolutional operations with 3x3 2D filters. The result was an output feature map with same size as the input, followed by max-pooling to down-sample the image. Downstream the convolutional blocks, the last three CNN layers were fully-connected. Particularly, following the last convolutional block, a fully-connected layer preceded the last two neurons of the output. This layer was responsible for assessing classification probabilities for IDH wild-type versus IDH-mutant GBMs. IDH mutation probability was calculated with the softmax algorithm for each sample, and categorical cross-entropy loss (CCEL) was chosen as objective function of our network with two output nodes. We employed 5-fold cross-validation, so that the performance measures average over five different splitting of training and testing data, keeping a minimum threshold of 5 IDH-mutant patients in the testing cohort. To demonstrate the individual predictive performance of different MRI sequences, T1-weighted-, T2-weighted-, FLAIR-, MPRAGE-, ADC-, rCBV-fed networks were trained for 500 epochs separately. Each epochs takes 7s to iterate on the entire training dataset, with a complex training time of 1h for each MRI sequence. As previously employed in a very similar study [27], Adam with a learning rate of 0.0001 was chosen as optimizer due to higher speed in reaching convergence in deep CNNs [28].

Model training was performed on NVIDIA-SMI GPU (CUDA Version, NVIDIA, Beijing, China). Our pipeline was written in Python3, using the Keras API Framework (https://keras.io/about/).

RESULTS

We selected 156 adult patients (mean age = 62 y, range = 35-83 y) with pathologically-proven diagnosis of GBM. Among these patients, 100 underwent IDH testing and were included in the study. Histopathological analysis demonstrated 83 IDH-wildtype (42 males, mean age 63 years) and 17 IDH-mutant GBM (11 males, mean age 56 years). All patients received postoperative focal radiotherapy plus concomitant daily temozolomide (TMZ), followed by adjuvant TMZ therapy, with the same treatment protocol.

IDH mutation prediction performance was averaged over five different splitting of training and testing data. The final model for T1-weighted, T2-weighted, FLAIR, MPRAGE, rCBV and ADC images achieved mean AUCs of 0.71, 0.63, 0.74, 0.62, 0.86, 0.45 respectively. ROC curves are reported in **Fig.3** for every MR sequence.

Accuracy, loss, sensitivity and specificity on the independent testing cohort were as follows: 77%, 1.4, 36%, 95% for T1-weighted sequence; 67%, 2.41, 48%, 75% for T2-weighted sequence; 77%, 1,98, 28%,95% for FLAIR sequence; 66%, 2.55, 43%, 74% for MPRAGE sequence; 83%, 0.64, 76%, 86% for rCBV sequence and 56%, 2.53, 14%, 73% for ADC. Predictive performances are reported in **Table 1**, with corresponding box-plots in **Fig.4**. In the supplementary materials, we reported learning plots fitted with an exponential function (1) for the testing cohort:

$$f(x) = a \times e^{-b \times x} + c \qquad (1)$$

DISCUSSION

This is the first study to attempt IDH status prediction in GBM. Our model achieved good prediction performance for IDH genotype, especially on rCBV maps (83% accuracy, 76% sensitivity, 86% specificity), showing comparable or superior results to other studies employing CNN architectures on gliomas [22,24,25,27].

Mutations of IDH coding gene lead to accumulation of D-2 Hydroxyglutarate (D-2HG), an oncogenic metabolite which may affect cellular differentiation [7,8,30]. This reflects on tumoral features such as cellularity, pattern of growth, vascularization, which present a radiologic correlate on MRI [31,32]. Even though the human eye may not capture these features with acceptable accuracy [5,17], modern artificial intelligence techniques can help overcoming these limitations. Radiomics achieved remarkable results in providing biomarkers for patient survival and tumor genetics [15]. For example Hsieh et al. achieved an accuracy of 85%, a sensitivity of 86%, and a specificity of 84% in predicting IDH mutation in GBM with texture features and a binary logistic regressor classifier [33]. Zhang et al. obtained 89% accuracy in predicting IDH mutation in HGG with a random forest classifier [34]. Both of these studies rely on supervised ML algorithms for the prediction tasks, which are not exempted from shortcomings: lack of parameter standardization may limit reproducibility and reliability of these models [35]. Our group recently investigated the performance of several ML classifier with different optimization parameters, highlighting how ensemble architectures show better results in clinical tasks prediction for GBM [unpublished data]. However, this standard radiomic workflow retains several steps with supervision constraints, such as feature extraction and selection [36,37].

In recent years, deep learning emerged as a promising technique to analyze imaging data [18]. Deep learning employs specific architectures named CNNs to achieve task prediction without human supervision [38]. Regarding our topic, Chang et al. predicted IDH mutation in a group of tumors including LGG and HGG by employing a 2D residual CNN (ResNet34) with 82.8% accuracy for training, 85.7% for testing and 83.6% for validation cohorts [24]. Starting from this architecture, Liang et al. implemented a Multimodal 3D DenseNet for IDH prediction achieving 84.6% accuracy, boosted to 91.4% when associated to transfer learning [25]. Other Authors [29] described a CNN for prediction of IDH mutation, MGMT methylation, and 1p/19q co-deletion from 256 brain MRIs from the Cancer Imaging Archives Dataset including LGG and HGG. They reported 94% accuracy for IDH, 92% for 1p/19q, and 83% for MGMT status. These studies rely on heterogeneous glioma populations [24,25,29], which may represent a bias when such results are applied to GMB alone. For example, since IDH-mutant GBMs are very rare, identification of IDH mutation may simply reflect tumor grade in an heterogeneous glioma group, especially when automatic predictive models are employed. Indeed, LGGs show typical radiologic features: less enhancement, less necrotic components and a diffused pattern of growth [5]. Differently, most

GBMs show peripheral enhancement and central necrosis on MRI, with surrounding admixture of infiltrative tumor and vasogenic edema [5] (**Fig.1**). Chang P. et al. applied principal component analysis to the final output layer of their CNN to visually display the highest-ranking features associated with IDH mutation [29]. As result, T1 post-contrast features demonstrated the highest rank: IDH-wildtypes showed thick and irregular enhancement or irregular, ill-defined, peripheral enhancement, while IDH-mutants displayed less enhancement and well-defined tumor margins. This result seem to capture the radiologic appearance of HGG vs LGG more than IDH-specific features [5].

Predicting IDH mutation in GBM is extremely important since conventional imaging features of IDH mutated tumors show considerable overlap with those of IDH-wild types [17] (**Fig.1**). Also, primary and secondary GBMs are generally indistinguishable at standard MRI, although secondary GBM harbors most of IDH mutations [5]. Our study proposes a new deep-learning model tailored to GBM, with high prediction performance for IDH status on MRI sequences. This has several advantages: the automatic pipeline of our model limits high-skilled programming, with easy-handling for further applications. Moreover our study is the first to explore the predictive power of advanced MRI sequences such as DWI and DSC perfusion.

The high IDH prediction accuracy obtained on perfusion images is consistent with previous studies [39,40]. Besides reflecting tumor grade [41], perfusion parameters showed promising correlation with patient survival, particularly rCBV [42]. In our experience, we achieved high predictivity of GBM IDH status with machine ML on rCBV data [unpublished data]. Kieckegereder et al. demonstrated that *IDH* mutation status is associated with a specific hypoxia/angiogenesis transcriptome signature through perfusion imaging [39]. Wu et al. extracted GBMs vascular habitats based on DSC perfusion, reporting that IDH mutation correlates to rCBV values of the low-angiogenic habitat from the enhancing tumor. These results reflect the known correlation between IDH mutation, hypoxia inducible factor and neoangiogenesis [43]. Our CNN architecture computes automated deep-features from input images, extracting semantic regional features including not only the tumor core but also the peritumoral edema as we choose to include the entire tumor from T2-ROI masks (see methods). A study from Chow et al. may support our choice, revealing that GBM induces vascular dysregulation in peritumoral regions, which are larger in IDH-wildtype than in IDH-mutant gliomas, helping to differentiate IDH genotypes [44]. One may argue that diffusion parameters were expected to achieve higher performance, since ADC correlates to tumor cellularity [44-46] and showed good results in predicting patient survival [47]. However, radio-genomic associations with IDH are poorly explored in GBM [16]. Also, diffusion parameters may fail to capture differences in a relatively homogeneous GBM group.

Our study has some limitations. First of all, the number of subjects may be considered small for deep-learning applications. We employed data augmentation (see methods) to partially compensate this issue. Due to the limited samples in our dataset, in order to prevent overfitting, we reduced the number of learning parameters generated by each layer. This prevented us from employing other architectures, such as ResNet, which generate a large number of parameters, making the training unfeasible. Also, our analysis relied on two cohorts only due to population size. We performed cross-validation to overcome the problem of randomization and the lack of a validation set. Differently from previous studies, we could not include patients from public datasets because of lack of perfusion images.

# CONCLUSIONS

Our deep-learning model achieved a maximal accuracy of 83% on rCBV images, in agreement with previous studies showing a correlation between IDH mutation and angiogenesis in gliomas. A GBM-specific model leads to several advantages to non-invasively estimate IDH mutation, disentangling the prediction from WHO grade.

# ACKNOWLEDGEMENTS

This study was supported by the grant 'Progetti di Ricerca Piccoli 2020' from La Sapienza University (Protocol ID: RP120172B9E252BD). Funding sources did not influence any phase of the present study.

FIGURES AND TABLES

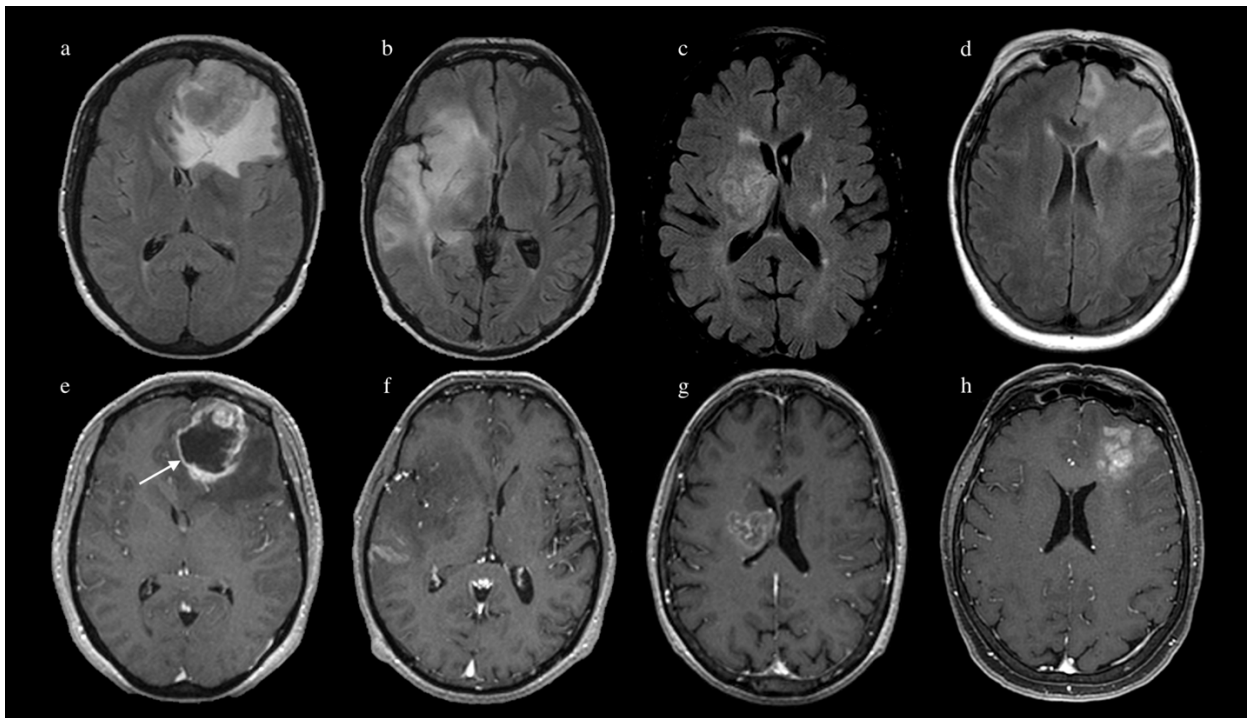

**Fig.1** FLAIR images (above) and post-contrast MPRAGE images (below) of four patients with GBM from our cohort. Patient 1 (a, e) presented an expansile left frontal lobe GBM with typical features of rim-enhancement and central necrosis (arrow on image e); pathology confirmed IDH-wildtype. Patient 2 (b, f) presented a diffuse non-enhancing infiltrative GBM of the right temporo-insular region; pathology confirmed IDH mutation. Patient 3 (c, g) and 4 (d, h) demonstrated 'borderline' features, not typical for any IDH status; pathology confirmed IDH mutation for patient 3 and IDH wild-type for patient 4

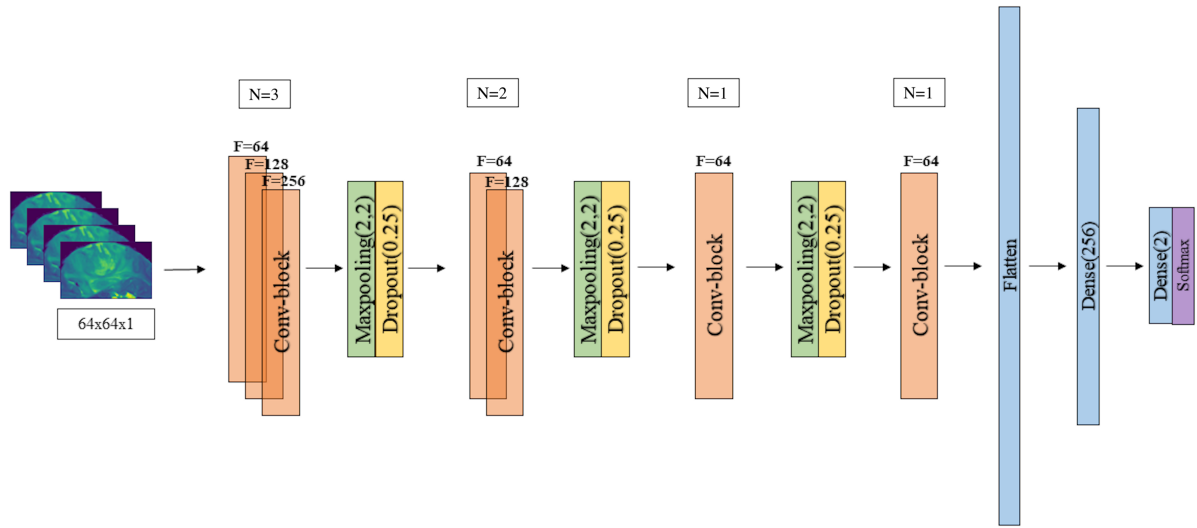

**Fig.2** Proposed CNN architecture to predict IDH status

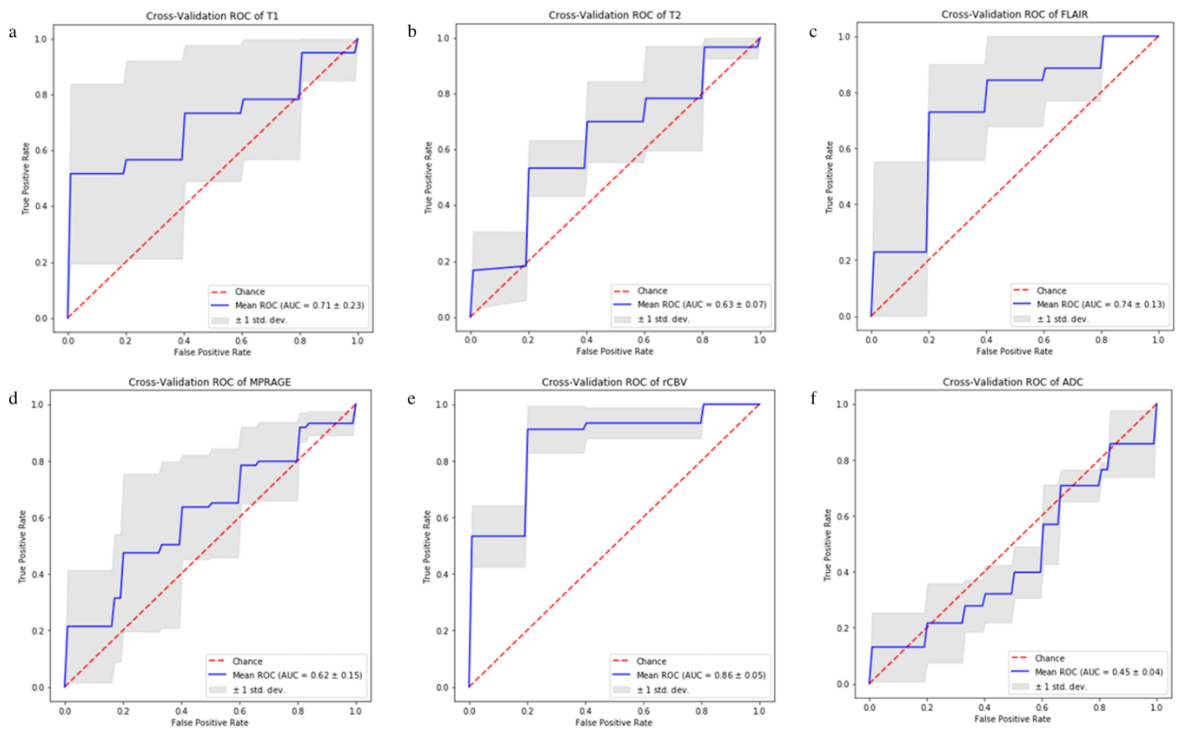

**Fig.3** ROC curve for testing set from k-fold cross validation (k=5) training on T1-weighted MRI sequence (a), T2-weighted MRI sequence (b), FLAIR MRI sequence (c), MPRAGE MRI sequence (d), rCBV MRI sequence (e), ADC MRI sequence (f)

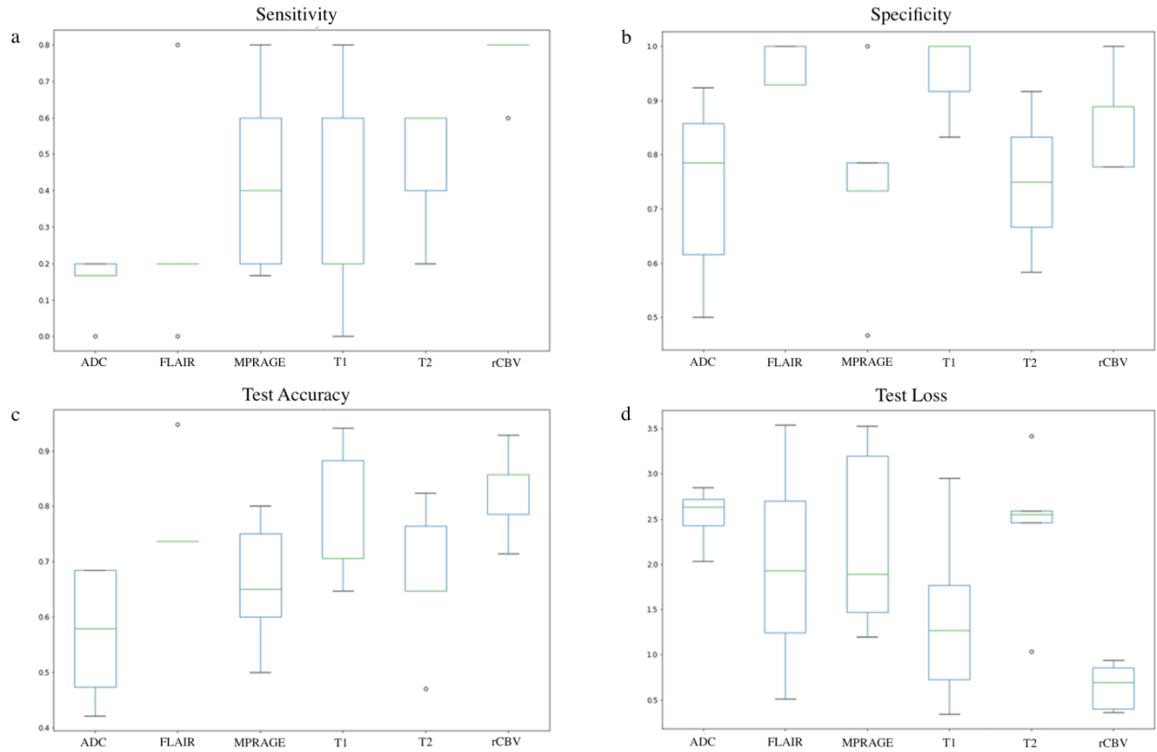

**Fig.4** Box-plot on k-fold cross validation (k=5) for sensitivity (a), specificity (b), test accuracy (c) and test loss (d) with all MRI sequences.

**Table 1** Predictive performance of each MR sequence indicated as mean±standard deviation

| Sequence | Test Acc | Test Loss | SN | SP | AUC |
|---|---|---|---|---|---|
| T1 | 0.77±0.11 | 1.4± 09 | 0.36±0.29 | 0.95±0.07 | 0.71±0.23 |
| T2 | 0.67±0.12 | 2.41±0.77 | 0.48±0.16 | 0.75±0.12 | 0.63±0.07 |
| FLAIR | 0.77±0.11 | 1.98±1.06 | 0.28±0.27 | 0.95±0.03 | 0.74±0.13 |
| MPRAGE | 0.66±0.1 | 2.55±0.93 | 0.43±0.24 | 0.74±0.17 | 0.62±0.15 |
| rCBV | **0.83±0.07** | **0.64±0.23** | **0.76±0.08** | **0.86±0.08** | **0.86±0.05** |
| ADC | 0.56±0.11 | 2.53±0.28 | 0.14±0.07 | 0.73±0.16 | 0.45±0.04 |

*Acc = accuracy; SN = sensitivity; SP = specificity; AUC = area under the curve